\documentstyle[11pt,a4]{article}

\def \square {\hbox{$\sqcup\!\!\!\!\sqcap$}} 
\newcommand{\be}{\begin{equation}}
\newcommand{\ee}{\end{equation}} 
\newcommand{\bea}{\begin{eqnarray}}
\newcommand{\eea}{\end{eqnarray}}

\catcode`\@=11 
\@addtoreset{equation}{section}
 
\catcode`\@=11

\begin{document}

\begin{titlepage}

\begin{flushright} 
{\tt 	FTUV/96-40\\ 
	IFIC/96-48\\ 
	hep-th/9606097}
 \end{flushright}

\bigskip%\vskip3mm

\begin{center}

{\bf{\LARGE Conformal and Nonconformal Symmetries\\ 
	in 2D Dilaton Gravity
\footnote{Work partially supported by the 
{\it Comisi\'on Interministerial de Ciencia y Tecnolog\'{\i}a}\/ 
and {\it DGICYT}.}}
}

\bigskip 

 J.~Cruz$^1$\footnote{\sc cruz@lie.uv.es},
 J.~Navarro-Salas$^1$\footnote{\sc jnavarro@lie.uv.es},
 M.~Navarro$^{2,3}$\footnote{\sc mnavarro@ugr.es} 
 and
 C.~F.~Talavera$^{1,4}$\footnote{\sc talavera@lie.uv.es}

\end{center}

\bigskip% 

\footnotesize
\begin{enumerate}	
\item Departamento de F\'{\i}sica Te\'orica and 
	IFIC, Centro Mixto Universidad de Valencia-CSIC.
	Facultad de F\'{\i}sica, Universidad de Valencia,	
        Burjassot-46100, Valencia, Spain. 
\item Instituto Carlos I de F\'\i sica Te\'orica y Computacional,
        Facultad  de  Ciencias, Universidad de Granada. 
        Campus de Fuentenueva, 18002, Granada, Spain. 
\item Instituto de Matem\'aticas y F\'\i sica Fundamental, 
        CSIC. Serrano 113-123, 28006 Madrid, Spain.
\item Departamento de Matem\'atica Aplicada.      
	E.T.S.I.I. Universidad Polit\'ecnica de Valencia.      
	Camino de Vera, 14. 46100-Valencia, Spain.                  
\end{enumerate}
\normalsize 

\bigskip%\vskip2mm 
%\centerline{\today}
\bigskip%\vskip2mm

%\newpage 

\begin{center}
			{\bf Abstract}
\end{center}

%\footnotesize 

We study finite-dimensional extra symmetries of generic 2D dilaton gravity models. 
Using a non-linear sigma model formulation we show that the unique theories 
admitting an extra (conformal) symmetry are the models with an exponential 
potential $V \propto e^{\beta\phi}$ ($ S ={1\over2\pi} \int d^2 x \sqrt{-g} 
[ R \phi + 4 \lambda^2 e^{\beta\phi} ]$), which include the model of Callan, 
Giddings, Harvey and Strominger (CGHS) as a particular though limiting 
($\beta=0$) case. These models give rise to black hole solutions with a 
mass-dependent temperature. The underlying extra symmetry can be 
maintained in a natural way in the one-loop effective action, thus implying 
the exact solubility of the semiclassical theory including back-reaction.
Moreover, we also introduce three different classes of (non-conformal)
transformations which are extra symmetries for generic 2D dilaton gravity models.
Special linear combinations of these transformations turn out to be the 
(conformal) symmetries of the CGHS and  $V \propto e^{\beta\phi}$ models.
We show that one of the non-conformal extra symmetries can be converted into a 
conformal one by means of adequate field redefinitions involving the metric 
and the derivatives of the dilaton. Finally, by expressing the 
Polyakov-Liouville effective action in terms of an invariant metric, we are 
able to provide semiclassical models which are also invariant.  This 
generalizes the solvable semiclassical model of Bose, Parker and Peleg (BPP)
for a generic 2D dilaton gravity model.

%\normalsize
\end{titlepage}

\newpage

\section{Introduction}

The physics of black holes provides an excellent arena on the
interphase between General Relativity and Quantum Mechanics. 
The fate of a black hole and the possible loss of quantum coherence are 
problems the final resolution of which requires to quantize the
gravitational degrees of freedom.
However, a consistent formulation of a quantum theory of gravity is still
lacking, and the semiclassical theory, that could serve as a first step to the
full quantized theory, is very difficult to solve.

A natural approach to bypass this situation is to consider simplified
models that keep the fundamental features of the four dimensional
theory but makes the analytic question more tractable. General
Relativity is trivial---up to global degrees of freedom---in lower
dimensions than four. However, some models derived as the S-wave sector of
the four dimensional theories are dynamically non-trivial and contain
formation of black holes by gravitational collapse. These models
involve the addition of a dilaton field to lower dimensional General
Relativity. 
The string-inspired model introduced by Callan, Giddings, Harvey and
Strominger (CGHS) \cite{CGHS} is one of the simplest toy models that
describe the formation of a black hole by gravitational collapse of massless
scalar fields. 
The quantum back-reaction is included at the one-loop level by adding the
Polyakov-Liouville term \cite{Polyakov} to the classical action.
In this simplified context, a special modification of the original
semiclassical CGHS model was introduced---the so-called RST model
\cite{RST}---making it possible to solve the semiclassical equations exactly
(see also \cite{Giddings}-\cite{Thorlacius}).

The key point in the formulation of the RST model is the addition of a
kinetic counterterm to the one-loop semiclassical CGHS action in such a way
that an extra symmetry is maintained at the quantum level thus
implying exact solubility.
More recently, a new quantum corrected version of the CGHS model was
introduced (the BPP model \cite{BPP}), which is also exactly solvable in the
semiclassical approximation including back-reaction (see also \cite{Cruz} for
the one-parameter class of models interpolating between the RST and BPP
models).
Unlike the RST model, in the BPP model the symmetry transformation is
exactly the same as in the classical one, and describes an evaporating black
hole with a non-flat end-state geometry.
Although both the RST and BPP models (among other models (see \cite{Kazama})
can be converted, through field redefinitions, into a Liouville model
possessing the infinite-dimensional conformal symmetry
(which guarantees the background independence in the sigma-model
formulation) the aim of this work is to explore the existence of finite-dimensional
extra symmetries for generic models of dilaton gravity.
The main consequence of the extra symmetries is that, as we will see, 
allow to define in a natural way a related semiclassical theory
invariant under the extra symmetry.

The 2D dilaton models would, nonetheless, be a highly more 
useful tool if  (at least part of) the developments 
which have been made with the CGHS theory would also be possible 
with more general models. In 
particular, for spherically symmetric Einstein gravity,
which is one of the most realistic models of 2D dilaton gravity.
In this paper this goal is pursued in two different though complementary 
directions. Firstly, we look for models which, alike the CGHS model, are  
invariant under an extra (conformal) symmetry. We shall show that, in the approach 
of the non-linear sigma models \cite{Bilal} these theories, 
when the kinetic term has been dropped out by an appropriate field 
redefinition, are restricted to 
have a potential of exponential form $V =4\lambda^2\hbox{e}^{\beta\phi}$ 
where $\phi$ is the dilaton field.
Secondly, we shall present three (model-dependent) transformations 
which are symmetries  for generic 2D dilaton gravity 
models with arbitrary potential $V=V(\phi)$. (This
includes spherically symmetric gravity, for which the
potential is $V\left(\phi\right)\propto {1\over\sqrt{\phi}}$).  
These symmetry transformations generalize that of the 
CGHS (BPP and RST) model in the sense that, although 
they are non-conformal for a 
generic potential and involve the space-time 
derivatives of the fields,  a linear combination of 
them is conformal and equals the extra conformal 
symmetry of the CGHS model in the particular 
case in which the potential is constant.  
Therefore, by following parallel lines to those which produce the 
semiclassical BPP model from the classical 
CGHS model, we are able to provide semiclassical models 
for a generic 2D dilaton gravity model which maintain an extra symmetry.

The organization of the paper is as follows.
In Sec.~2 we briefly review the semiclassical theory of the CGHS model, 
and present a simple procedure to construct a family of quantum corrected
actions maintaining the classical free field equation.
In Sec.~3 we find out the exponential models 
as the only models with an extra (conformal) symmetry. 
Sec.~4 is devoted to analyze the main semiclassical aspects of these models.  
In Sec.~5 we present the main goal of the paper.
We introduce three new extra symmetries for generic 2D dilaton
gravity. 
We work out an invariant metric for a particular (non-conformal) symmetry
and construct an invariant semiclassical action for a generic theory. 
In Sec.~6 we state our conclusions.

\section{Semiclassical theory of the CGHS model}

A particularly simple model for black hole physics \cite{CGHS}\ is given by
the classical action
\be
S_0 = {1\over2\pi} \int d^2 x \sqrt{-g}
	\left[ e^{-2\phi} (R + 4 (\nabla\phi)^2 + 4\lambda^2 ) -
	{1\over2} \sum_{i=1}^N (\nabla f_i)^2 \right]
\>,
\label{xxviii}
\ee
where $\phi$ is the dilaton field and $f_i$ are a set of $N$ massless
scalar fields. The solubility of the theory is based on the existence of a
free field.
In conformal gauge (i.e, $ds^2=-e^{2\rho}dx^+dx^-$), one has the
free field equation
\be
\partial_+ \partial_- (\rho -
\phi) =0\>.\label{z}
\ee
This is a consequence of the following symmetry
\be
\delta \phi = \delta \rho = \epsilon \, e^{2\phi}
\>. \label{cosai}
\ee
A simple way to see the invariance of the classical action under the above
transformations is to perform the following redefinition of variables
\bea
\bar{g}_{\mu\nu} &=& g_{\mu\nu} e^{-2\phi} \>, \nonumber \\
\bar{\phi} &=& e^{-2\phi} \>. 
\eea
In terms of $\bar{g}_{\mu\nu}$ and $\bar{\phi}$, the classical CGHS action (\ref{xxviii})
takes the reduced form
\be
S_0 (\bar{g}, \bar{\phi}) = {1\over2\pi} \int d^2 x \sqrt{-\bar{g}} 
\left[ \bar{R} \bar{\phi} + 4 \lambda^2 -{1\over2}\sum_{i=1}^N\left(\nabla f_i\right)^2\right]
\>. \label{cosaii}
\ee
The transformation (\ref{cosai}), when written in terms of the new fields,
adopts the trivial form
\bea
\delta\bar{g}_{\mu\nu} &=& 0 \>, \nonumber \\
\delta\bar{\phi} &=& \epsilon \>, \label{cosaiia}
\eea
and the action (\ref{cosaii}) is invariant because $\sqrt{-g} R$ is a total
derivative in two dimensions.
To account for the back-reaction effect one has to 
add the one-loop Polyakov-Liouville term
\be
S_P=-{N\over96\pi}\int d^2 x\sqrt{-g}R\square^{-1}R\>.\label{r1}
\ee
This term breaks the symmetry (\ref{cosai}) 
although the RST counterterm 
\be
S_{RST}=-{N\over96\pi}\int d^2x\sqrt{-g}2\phi R\>,\label{r2}
\ee
reestablishes it with the 
following correction:
\be
\delta\phi =\delta\rho = \epsilon { e^{2\phi} \over \left( 1 - {k\over4}
e^{2\phi}\right) }
\>.
\ee
Therefore, the conserved current $\partial_+ \partial_- (\rho - \phi)$ is
maintained at one-loop level.

However, the easiest way to construct a semiclassical action, invariant
under the transformations (\ref{cosai}), is to consider the
Polyakov-Liouville term with respect to the invariant metric $\bar{g}_{\mu\nu}$.
Going back to the fields ($g_{\mu\nu}$, $\phi$), the effective action
decomposes into
\be
S_P (\bar{g}) =S_P(g) + S_{BPP} (g,\phi)
\>, \label{bppi}
\ee
where the local term $S_{BPP}(g,\phi)$ turns out to be the counterterm of the
Bose-Parker-Peleg model \cite{BPP}:
\be
S_{BPP} (g,\phi) = {N\over24\pi} \int d^2 x \sqrt{-g} 
\left( (\nabla\phi)^2 - \phi R \right)
\>. \label{bppii}
\ee
This also provides a way to select a particular metric in the functional integral of conformal matter fields.
The metric is chosen to be invariant with respect to the symmetry of the classical dilaton-gravity action.

At this point it is interesting to comment that, despite of the fact that
the classical equations imply the vanishing of the scalar curvature $\bar{R}$,
and therefore the trace anomaly, the black holes do radiate.
Indeed, the BPP model describes the formation and evaporation of a black
hole producing a non-trivial remnant geometry.
Moreover, the BPP model has also emerged as the semiclassical limit of the
non-perturbative approach of Ref. \cite{Mikovic}.

It is easy to see now that one can construct a general class of one-loop
models preserving the current conservation equation $\partial_+ \partial_-
\bar{\rho} \equiv \partial_+ \partial_- (\rho-\phi) =0$.
If we introduce a new field (where $G$ is an arbitrary function)
\be
\tilde{\phi} =\bar{\phi} + {N\over12} G(\bar{\phi})
\>,
\ee
and modify the classical action (\ref{cosaii}) by
\be
S_0\left(\bar g,\bar\phi\right) 
\quad\longrightarrow\quad
S_0 \left( \bar g,\bar\phi+{N\over12} G\left(\bar\phi\right) \right)
\>, \label{gvi}
\ee
it is clear that the new action is invariant under the transformation
\bea
\delta \bar g_{\mu\nu}&=&0 \>, \nonumber\\
\delta \tilde\phi&=&\epsilon \>.\label{gvii}
\eea
Returning now to the primitive  fields ($g_{\mu\nu}$, $\phi$), the action
\be
S_0 (\bar{g}, \tilde{\phi}) + S_P(\bar{g})
\ee
turns out to be
\bea
&&S_0 (g,\phi)
+ S_P(g) \label{cosax} \\
&&+ {N \over 24\pi} \int d^2 x \sqrt{-g}
\left[ (\nabla\phi)^2 - 2 g^{\mu\nu} \nabla_\mu F(\phi) \nabla_\nu \phi
+ F(\phi) R - \phi R \right]
\>, \nonumber
\eea
where $F(\phi) = G\left(\bar\phi\left(\phi\right)\right)$, and the transformations
(\ref{gvii}) take the form
\bea
\delta g_{\mu\nu} &=&
 \epsilon { 2 e^{2\phi}
 \over 1-{N\over24}F^{\prime}\left(\phi\right)e^{2\phi}}g_{\mu\nu}
 \>, \label{gviii}\\
\delta\phi &=& \epsilon{e^{2\phi}\over 1-{N\over24}F^{\prime}
\left(\phi\right)e^{2\phi}}
\>.\label{gix}
\eea
The above family of 2D dilaton gravity models, parametrized by the arbitrary
function $F$, was obtained in Ref. \cite{Cruz} (see also \cite{Michaud}) by suitable field redefinitions
from the classical CGHS action corrected with a (non-covariant)
Polyakov-type term, invariant under Weyl transformations
\cite{Jackiw} \cite{Navarro}.
The RST model can be recovered with the choice $F(\phi) = {1\over2}\phi$.
In general, the field redefinitions
\bea
\Omega &=& \sqrt{k} F(\phi) + {e^{-2\phi} \over \sqrt{k}} \>, \nonumber \\
\chi &=& \sqrt{k} \, \rho + \sqrt{k} \left( F(\phi) - \phi \right) +
{e^{-2\phi}\over\sqrt{k}} \>, 
\eea
convert the models (\ref{cosax}) into a Liouville theory 
\be
{1\over\pi}\int d^2x\left[-\partial_+\chi\partial_-\chi+\partial_+\Omega\partial_-\Omega+
\lambda^2 e^{{2\over\sqrt{\chi}}\left(\chi-\Omega\right)}+{1\over2}\sum_{i=1}^N
\partial_+f_i\partial_-f_i\right]\>.\label{r4}
\ee
We have to mention now that the Polyakov-Liouville term coupled to the invariant metric
seems to be the unique semiclassical term which, for a generic theory (see for instance
the exponential model of section 4),
maintains the extra symmetry.

\section{Extra (conformal) symmetries in 2D dilaton gravity}
\label{sec:conformal}

A natural way to analyze the existence of extra symmetries in generic dilaton
gravity theories, generalizing the powerful symmetry of the CGHS model, is
based on the non-linear sigma model formulation of these theories
. Let us now review briefly the results of Ref.~\cite{Kazama}.

The expression of a non-linear sigma model action associated with
a wide class of 2D dilaton gravity models is
\be
S = \int d^2x\sqrt{-\hat g} 
\left[\hat{g}^{\mu\nu}
\partial_{\mu}X^i G_{ij}\left(X\right) \partial_{\nu}X^j 
+ Q\left(X\right) \hat{R}
+ \Lambda e^{W\left(X\right)}\right]
\>,\label{aix}
\ee
where $\Lambda$ is a constant, $\hat{g}_{\mu\nu}$ is a reference metric, and
$G_{ij}$ is a metric depending on the target
space coordinates $X^i (x)$ ($i=1,2$) ($X^1$ and $X^2$ stand for the dilaton
field and the conformal factor respectively).
If we consider now a flat reference metric 
$\hat{g}_{\mu\nu} = \eta_{\mu\nu}$, 
(\ref{aix}) reduces to
\be
 S_{flat}=\int d^2x \left[\partial_{\mu}X^i \, G_{ij}\, \partial^{\mu}X^j
 +\Lambda e^{W}\right]
\>.\label{ax}
\ee
 The change of the Lagrangian under a general variation
 $\delta X^i$ is
 \bea
 \delta {\cal L}&=& 2\partial^{\mu}\left(\partial_{\mu}X^iG_{ij}\delta X^j\right)
 \nonumber \\
&&  -2\left( \square X^k+\Gamma^k_{ij}\partial_{\mu}X^i\partial^{\mu}
 X^j \right) G_{kl}\delta X^l
 + \Lambda W,_k\delta X^k e^W\>,\label{axi}
 \eea
 where $\Gamma^k_{ij}$ is referred to the metric $G_{ij}$, $\square$ is
 the world sheet Laplacian and $W,_{k}$ denotes derivative with 
 respect to the target space coordinate $X^k$.
 It is easy to check that there exists a symmetry if the 
 variation $\delta X^i$ satisfies 
 \bea
W,_k\delta X^k 		&=& 0\>,\label{axii}\\
\nabla_i\delta X_j 	&=& \partial_i\delta X_j -\Gamma^k_{ij}\delta X_k=0
 \>.\label{axiii}
 \eea
 Equation (\ref{axii}) can be solved to give
 \be
 \delta X^k={\epsilon^{kl}\over\sqrt{-|G|}}W,_l\>,\label{axiv}
 \ee
 being $|G|=\det G_{ij}$, and $\epsilon^{kl}$ is the Levi-Civita symbol.
 Substitution of (\ref{axiv}) into equation (\ref{axiii})
 yields the condition
 \be
 \nabla_i\nabla_j W=0
 \>.\label{axv}
 \ee
An immediate consequence of this symmetry is that it allows to construct a
free field $F(x)$. $F(x)$ is associated with the corresponding 
Noether current $j^\mu$:
\be
j_\mu = \partial_\mu X^i G_{ij} {\epsilon^{jk} \over \sqrt{-|G|}} W_{,k}
\ee
by the simple relation
\be
j_\mu = - \partial_\mu F \>,
\ee
where the condition (\ref{axv}) guarantees the existence of the function
$F(x)$. Moreover, the field $F(x)$ is orthogonal to the field $W(x)$, which
satisfies the Liouville equation.

\bigskip 

Let us return now to the generally covariant expression of a generic 2D
dilaton gravity model,
\be
S\left(g,\phi\right) = {1\over2\pi} \int d^2x\sqrt{-g}
\left[ D(\phi) R + {1\over2} \left(\nabla\phi\right)^2 + V(\phi) \right]
\>,\label{ai}
\ee
where $D(\phi)$ and $V(\phi)$ are arbitrary functions of the dilaton field.
It is not difficult to see that one can eliminate the kinetic term for the
dilaton by a conformal redefinition of the fields \cite{Banks} \cite{Gegenberg}.
Therefore, one can study a generic 2D dilaton gravity model form the action
\be
S \left(g,\phi\right) = {1\over2\pi} \int d^2x\sqrt{-g}
\left(R\phi + V (\phi)\right)\>,\label{av}
\ee
For the CGHS model we have $V = 4\lambda^2$, and the
Jackiw-Teitelboim
 model \cite{JT} and spherically symmetric gravity \cite{Strominger} correspond to
$V = \Lambda \phi$ and 
$V = \lambda^2 / \sqrt{2\phi}$, respectively.

Our aim now is to classify the extra (conformal) symmetries of the 2D dilaton
gravity models according to the criteria (\ref{axv}).
This can be achieved immediately if one consider the reduced form (\ref{av})
of generic dilaton gravity in conformal gauge,
\be
 S ={1\over2\pi} \int d^2x 
 \left(-4\partial_+\phi \partial_-\rho
 + {1\over2} V(\phi) e^{2\rho}\right)
 \>.\label{axvi}
 \ee
 This action fits the general expression (\ref{ax}), where
 \begin{displaymath}
 G_{ij}=\left[\begin{array}{cc}0&1\\1&0\end{array}\right]\>,\label{axix}
 \end{displaymath}
and 
\be
 W(X) = 2\rho + {1\over2}\log V (\phi) 
\>. \label{axviii}
\ee
As the components of $G_{ij}$ are constants, the condition (\ref{axv}) 
becomes
\be
 {d^2\log V\left(\phi\right)\over d\phi^2}=0
\>.\label{axx}  
\ee
Thus the general solution is $V(\phi) = 4\lambda^2 e^{\beta\phi}$, where
$\lambda$ and $\beta$ are constants.
The CGHS model is recovered for $\beta=0$.
In this case the conformal symmetry $\delta X^i$ is (\ref{cosaiia}) 
and the free field $F = \rho$.

\section{The $V \propto e^{\beta\phi}$ model}

The CGHS model has provided an useful arena to describe back-reaction
effects in the black hole evaporation process.
The crucial property of the model is the existence of a powerful symmetry
that can be maintained at the one-loop level and, therefore, guarantees the
solubility of the semiclassical theory.
The models with an exponential potential $V=4\lambda^2 e^{\beta\phi}$
possesses also a similar conformal symmetry which can be exploited to produce
an exactly solvable semiclassical model.

\subsection{Classical equations}

In conformal gauge, the equations of motion of the model
\be
S_\beta = {1\over2\pi} \int \sqrt{-g} \left[ R \phi + 4\lambda^2
e^{\beta\phi} - {1\over2} \sum_{i=1}^N \left(\nabla f_i\right)^2 \right]
\label{betai}
\ee
are equivalent to
\bea
\partial_+ \partial_- \left(\rho - {\beta\over2} \phi \right)&=&0 \>,
\label{betaiia}\\
\partial_+ \partial_- \left(\rho + {\beta\over2} \phi \right) &=& -\lambda^2
\beta e^{2\left(\rho + {\beta\over2} \phi\right)} \>, \label{betaiib} \\
-\partial_{\pm}^2 \phi + 2 \partial_\pm \rho \, \partial_\pm \phi &=&
T^f_{\pm\pm} \>. \label{betaiic}
\eea
The field $\rho - {\beta\over2 } \phi$ is a free field and 
$\rho + {\beta\over2} \phi$ verifies the Liouville equation.
In Kruskal gauge $\left(\rho - {\beta\over2}\phi = 0\right)$, the equation (\ref{betaiib})
is
\be
\partial_+ \partial_- (2\rho) = - \lambda^2 \beta e^{4\rho}
\>, \label{betaiii}
\ee
the general solution of which can be written as
\be
\rho = {1\over4} \log {\partial_+ F \partial_- G \over \left( 1 + \lambda^2
\beta F G \right)^2 }
\>, \label{betaiv}
\ee
with $F$ ($G$) an arbitrary function of $x^+$ ($x^-$).
Using the above expression, the constrained equations (\ref{betaiic}) become
Ricatti differential equations with respect to $\log \partial_+ F$ and 
$\log\partial_- G$:
\bea
{1\over4} \left(\partial_+ \log \partial_+ F \right)^2 -
{1\over2} \partial_+^2 \log \partial_+ F &=& \beta T^f_{++} \>,
\label{betava} \\
{1\over4} \left(\partial_- \log \partial_- G \right)^2 -
{1\over2} \partial_-^2 \log \partial_- G &=& \beta T^f_{--} \>.
\label{betavb}
\eea

In the absence of matter, the general solution is
\be
e^{2\rho} = {1\over C x^+ x^- + A x^+ + B x^- + D} \>, \label{betavi}
\ee
where $A$, $B$, $C$ and $D$ are arbitrary constants with the restriction
\be
A B - C D = - \lambda^2 \beta
\>. \label{betavii}
\ee

If $C < 0$, the solution (\ref{betavi}) is similar to the CGHS black hole
solution.
The event horizon is located at $x^- = -A/C$, and the curvature singularity
is space-like for $\lambda^2\beta < 0$.
For a positive value of $\lambda^2\beta$ we have a naked singularity.
On the other hand, if $C > 0$, the curvature singularity is also naked for
$\lambda^2\beta > 0$, but the metric has not the appropriate signature.
When $C=0$, 
an elementary change of coordinates brings the metric to the
form
\be
ds^2 = {1\over\sqrt{|\lambda^2\beta}|} { - dt^2 + dx^2 \over 2x}
\>. \label{betaviii}
\ee
Although this space-time is geodesically complete, one can check by solving 
(4.7)(4.8) for a collapsing shock-wave of
matter (or by matching static solutions) that it produces a naked singularity
for $\lambda^2\beta > 0$.
Therefore, assuming a cosmic censorship conjecture we shall restrict the model 
to $\lambda^2\beta < 0$ and consider only the
elementary solutions with $C < 0$.

In analogy with the CGHS model, one can introduce asymptotically flat
coordinates $\{\sigma^\pm\}$ by
\be
\sqrt{|C|} \, x^\pm = \pm e^{\pm\sqrt{|C|} \sigma^\pm}
\>. \label{betax}
\ee
The resulting metric, for $A=0=B$, is then
\be
ds^2 = { - d\sigma^+ d\sigma^- \over 1 + {\lambda^2 \beta\over
C}e^{-2\sqrt{|C|}\sigma}}
\>. \label{betaxi}
\ee
At this point, one can use the Euclidean continuation $t \to i\tau$ of the
black hole metric to obtain the corresponding Hawking temperature.
Introducing a new spatial coordinate $R^2$ defined by (the horizon is
located at $R=0$)
\be
1 + {\lambda^2\beta\over C} e^{-2\sqrt{|C|}\sigma} = {1 \over R^2 |C|} \>,
\label{betaxii}
\ee
the Euclidean line element becomes
\be
ds_E^2 = R^2 d\left( \sqrt{|C|}\tau  \right)^2 
+ { dR^2 \over \left( 1 - |C| R^2 \right)^2} 
\>. \label{betaxiii}
\ee
The Hawking temperature, which is the inverse of the period of $\tau$, is
\be
T_H = {\sqrt{|C|} \over 2\pi}
\>. \label{betaxiv}
\ee
The static black hole solution (4.14) has the Killing vector
$k^{\mu}=\left({\partial\over\partial t}\right)^{\mu}=(1,0)$.
The associated Noether charge $Q$ is the mass of the
black hole and is given by
\be
Q={1\over2\pi}\epsilon^{\mu\nu}\left[2k_{\mu}\nabla_{\nu}\phi+\phi\nabla_{\mu}k_{\nu}\right]|_{\sigma=+\infty}
\>. \label{Wald}
\ee
In order to calculate $Q$ it is useful to introduce a new spatial
coordinate $x$ related to $\sigma$ by
\be
1+{\lambda^2\beta\over C}e^{-2\sqrt{|C|}\sigma}={1\over 1-e^{-2\sqrt{|C|}
x}}\>.\label{Wald1}
\ee
In the coordinates $\left(t,x\right)$
the metric takes the Schwarzschild-type form
\be
ds^2=-\left(1-e^{-2\sqrt{|C|}x}\right)dt^2+{1\over 1-e^{-2\sqrt{|C|}x}}
dx^2
\>,\label{Wald2}
\ee
and the dilaton is
\be
\phi=-{2\over\beta}\sqrt{|C|}x-{1\over\beta}\log{\lambda^2\beta\over C}
\>.\label{Wald3}
\ee
The charge $Q$ is now easily calculated
\be
Q=-{1\over2\pi}\left(k_0\nabla_1\phi-\phi\nabla_1 k_0\right)|_{x=+\infty}
={2\over\beta\pi}\sqrt{|C|}
\>.\label{Wald4}
\ee
This result can also be obtained by evaluating the ADM mass of the
 solution (\ref{betaxi}). (See \cite{Banks}). 
Therefore, in sharp contrast with the CGHS model, the Hawking temperature is proportional
to the black hole mass 
\be
T_H={\beta\over4}M\>.\label{temperature}
\ee
The existence of 2D black holes whose mass is proportional to their
temperature was noticed in \cite{Mann1}.
The black hole solutions (4.14) also appear, through a different
coupling of matter to 2D gravity, in \cite{Mann2}.

Finally, we would like to point out that, in parallel with the CGHS model, 
the present model can 
also be brought, through a conformal redefinition of the metric, to a form
for which the dynamics implies the vanishing of the scalar curvature.
In terms of the new metric $\tilde g_{\mu\nu}=e^{-\beta\phi}g_{\mu\nu}$, 
the action takes the form
\be
S_{\beta}={1\over2\pi}\int d^2x\sqrt{-\tilde g}\left(\tilde R\phi+\beta\left(\tilde\nabla\phi\right)^2+
4\lambda^2e^{2\beta\phi}\right)\>,\label{betaxv}
\ee
and the trace of the constrained equations is equivalent to the equation
\be
\tilde R=0\>.\label{betaxvi}\ee

\subsection{Semiclassical theory}

We shall now consider the semiclassical theory. 
The special symmetry of the model (\ref{betai}) is
\be
g_{\mu\nu}\longrightarrow e^{-\beta\epsilon}g_{\mu\nu}\>,\label{betaxviii}
\ee\be
\phi\longrightarrow \phi+\epsilon\>,\label{betaxix}
\ee
Following the procedure of Section 2 we can construct an one-loop corrected theory invariant under 
the transformations (\ref{betaxviii}) (\ref{betaxix}) by adding to (\ref{betai}) a Polyakov-Liouville term
$S_P\left(\bar g\right)$ respect to the invariant metric
\be
\bar g_{\mu\nu}=e^{\beta\phi}g_{\mu\nu}\>.\label{betaxx}
\ee
In terms of the fields $\left(\bar g_{\mu\nu},\phi\right)$, 
the semiclassical action is
\bea
&& {1\over2\pi}\int d^2x \sqrt{-\bar g}\left(\phi R\left(\bar g\right) 
- \beta\left(\bar \nabla\phi\right)^2
+ 4\lambda^2
- {1\over2}\sum_{i=1}^N\left(\bar\nabla f_i\right)^2\right) \nonumber\\
&& \qquad -{N\over96\pi}\int d^2x\sqrt{-\bar g}R\left(\bar g\right)\bar {\square}^{-1}
R\left(\bar g\right) \>,\label{betaxxi}
\eea
and, in conformal gauge, the equations of motion are:
\bea
\partial_+\partial_-\left(\bar\rho-\beta\phi\right) &=& 0
\>,\label{betaxxii} \\
\partial_+\partial_-\phi+\lambda^2e^{2\bar\rho} 
+ {N\over12}\partial_+\partial_-\bar\rho &=& 0
\>,\label{betaxxiii} \\
-\partial^2_{\pm}\phi
+ 2\partial_{\pm}\bar\rho\partial_{\pm}\phi
- \beta\left(\partial_{\pm}\phi\right)^2 \qquad &&
\nonumber \\
+ {N\over12}\left[\left(\partial_{\pm}\bar\rho\right)^2 
	- \partial^2_{\pm}\bar\rho+t_{\pm}\left(x^{\pm}\right)
	\right] &=& T_{\pm\pm}^f
\>.\label{betaxxiv}
\eea
Going back to the physical metric 
$\rho=\bar\rho-{\beta\over2}\phi$, and in Kruskal gauge ($\bar\rho-\beta\phi=\rho-{\beta\over2}\phi=0$),
the semiclassical equations can be written in the form
\be
\partial_+\partial_- (2\rho) = 
	{-\lambda^2\beta\over 1+{N\beta\over12}}e^{4\rho}\>,\label{betaxxv}
\ee
\be
e^{2\rho}\partial_{\pm}^2e^{-2\rho} = 
     {\beta\over1+{N\beta\over12}}\left(T^f_{\pm\pm}-{N\over12}t_{\pm}\right)
\>.\label{betaxxvi}
\ee
So, in Kruskal gauge, the one loop corrected equations are the same as 
the classical ones (\ref{betaiia})-(\ref{betaiic}), up to
a quantum shift for the $\beta$ parameter 
\be
\beta\longrightarrow{\beta\over1+{N\beta\over12}}\>,\label{betaxxvii}
\ee
and the addition of the non-local terms $t_{\pm}$ coming from the 
Polyakov-Liouville action.

Let us now consider static black hole solutions. 
With the choice 
$T^f_{\pm\pm}=0$ and $t_{\pm}=0$, 
we recover the solutions (\ref{betavi})-(\ref{betavii})
with the quantum corrected $\beta$ parameter (\ref{betaxxvii}).
In asymptotically flat coordinates $\{\sigma^{\pm}\}$ (\ref{betax}), 
the energy flux at infinity gives rise to a constant thermal value.
Using the anomalous transformation law of $<T^f_{\pm\pm}>$ we arrive at
\be
<T^f_{\pm\pm}\left(\sigma^{\pm}\right)>=-{N\over12}t_{\pm}\left(\sigma^{\pm}\right)={N\over48}|C|\>,
\label{betaxxviii}
\ee
corresponding to the Hawking temperature $T_H={\sqrt{|C|}\over2\pi}$.
The solution describes then a black hole in thermal equilibrium at temperature $T_H={\sqrt{|C|}\over2\pi}$.
Due to the non-static character of the solution with zero mass it is unclear how can we study
the evaporation process of a black hole formed by gravitational collapse.
 It could be considered instead the evolution of a black hole in thermal equilibrium
when it absorbs an incoming shock wave.
In the CGHS model the black hole remains static since the temperature is 
independent of the mass.
However, for the exponential model the shift in the mass due to the infalling 
shock wave increases the temperature and produces evaporation. 
These questions are out of the aim of the present work and will be considered 
elsewhere.

\section{Extra (non-conformal) symmetries in 2D dilaton gravity models} 
 
	In this section we shall consider the symmetries in 2D dilaton 
gravity models from a different standpoint.   
We shall show that suitable modifications of the  
conformal symmetries we have dealt with up to now 
produce transformations of the fields which are symmetries 
of all the  models  whose action can be brought to the form:   
\be 
{\cal L} = \frac1{2\pi}\sqrt{-g}\left[ R\phi + 
V(\phi)\right]
\label{lag}
\>.
\ee
In this way, we shall be able to generalize the  
procedure described in 
Section 2 ---which produces the BPP and RST models 
from the CGHS model and which requires the construction of 
a metric which is invariant under the symmetry---  
to a generic model of 2D dilaton gravity, thus 
generalizing the BPP and RST models. 

To find out these symmetries 
our strategy consists in finding the
generalized  conserved currents firstly. Then, by applying 
a very useful (sort of reciprocal) version  of the Noether 
theorem which is presented next, we shall 
find out the symmetries these (Noether)
currents are associated with. 

For any Lagrangian ${\cal L}={\cal L} \left(\Psi^a\right)$  
and arbitrary transformations of the fields $\delta\Psi^a$, 
we have:  
\be
\delta {\cal L}=
\left(E-L\right)_a\delta\Psi^a-\nabla_{\mu}s^{\mu}
\>,\label{hiii}
\ee
where $\left(E-L\right)_a=0$ are the Euler-Lagrange 
equations of motion for the fields $\Psi^a$,   
and $\nabla_{\mu}s^{\mu}$ is a 
total derivative term which appears 
due to the ``integrations by parts" which are generally 
required to produce the equations of motion. 
 
Let now $j^\mu$ be a current, which is made from 
the fields $\Psi^a$ and which is conserved on-shell: 
$\nabla_\mu {j^\mu}_{|_{\scriptstyle sol}}=0$.   
It is easy to see then that a transformation $\delta\Psi^a$ 
is the Noether symmetry associated to $j^{\mu}$ iff, 
without using any of the equations of motion, 
the following equality holds as an identity:   
\be
\left(E-L\right)_a\delta\Psi^a=\nabla_{\mu}j^{\mu}\>.
\label{hiv}
\ee
In general, and due to semi-invariance,  
the current $j^\mu$ will not be equal to $s^\mu$. 

For the Lagrangian in eq. (\ref{lag}) we have: 
\bea 
\delta {\cal L} &=& \frac{\sqrt{-g}}{2\pi}
\left\{\vphantom{{1\over2}}
\left[ R + V'(\phi)\right]\delta \phi\right. \nonumber \\
&&+\left[\nabla_\mu\nabla_\nu\phi+{1\over2} g_{\mu\nu} V\left(\phi\right)
-g_{\mu\nu}\nabla^2\phi\right]\delta g^{\mu\nu}\label{deltalag}\\
&&+ \left.\nabla_\alpha\left[-\phi\left(g^{\mu\nu}\nabla^\alpha
\delta g_{\mu\nu}   - g^{\alpha\mu}\nabla^\nu\delta g_{\mu\nu}\right))
   + \nabla^\alpha\phi \, g_{\mu\nu}\delta g^{\mu\nu} -
   \nabla_\nu\phi\, \delta g^{\nu\alpha}\right]\right\}\nonumber
\>.
\eea
Taking into account that $G_{\mu\nu}=0$ implies $\square\phi=V$ the 
equations of motion are equivalent to 
\bea 
R+ V'(\phi)&=&0	\>,	\label{r=0}\\
\nabla_\mu\nabla_\nu\phi 
-\frac12g_{\mu\nu}V(\phi)&=&0 \>, \label{nablamunu=0}
\eea 
 and then it is not difficult to check that the following currents  
 are conserved: 
\bea
j_{1}^{\mu}&=&{\nabla^{\mu}\phi\over\left(\nabla\phi\right)^2}
\>,\label{hi}
 \\
 j_2^{\mu}&=&j_R^{\mu}+V{\nabla^{\mu}\phi\over\left(\nabla\phi\right)^2}
 \>,\label{hii}
 \eea
 where $\nabla_{\mu}j^{\mu}_R=R$. 
 
 Now we can make use of the Noether theorem to show 
 that these currents are, in fact,
 Noether currents associated with symmetry 
transformations of the theory. 
The transformations which satisfy 
(\ref{hiv}) for the currents $j_1$ and $j_2$
are, respectively,
\bea
\delta_1\phi=0 
&,& 
\delta_1 g_{\mu\nu}=
\epsilon_1\left({g_{\mu\nu}\over\left(\nabla\phi\right)^2}
-2{\nabla_{\mu}\phi\nabla_{\nu}\phi\over\left(\nabla\phi\right)^4}\right)
\>, \label{hv}\\
\delta_2\phi = \epsilon_2
&,&
\delta_2 g_{\mu\nu}=\epsilon_2 V\left({g_{\mu\nu}\over\left(\nabla\phi
\right)^2}-2{\nabla_{\mu}\phi\nabla_{\nu}\phi
	\over\left(\nabla\phi\right)^4}\right) 
\>.\label{hvii}
\eea 

Though neither of these variations, $\delta_1$ or $\delta_2$, reproduce the 
conformal symmetry of the CGHS model when the potential $V$ is constant, 
$V=4\lambda^2$, it is easy to show that a linear combination of them do: 
\be
\delta=\delta_2-4\lambda^2\delta_1 
\>.\label{delta}
\ee
Therefore the transformation $\delta$ must be regarded as a symmetry 
which generalizes that of the CGHS model. 

Observe that both $\delta_1$ and $\delta_2$ are area-preserving (i.e.,  
$\delta_{1,2}\sqrt{-g}=
-{1\over2} \sqrt{-g} \, g_{\mu\nu} \break \delta_{1,2} g^{\mu\nu}=0$).
It is also interesting to remark at this point that  
when the fields are taken to be on-shell   
the symmetry transformation $\delta_2$ can be identified with a 
diffeomorphism, with infinitesimal space-time vector field 
$f^\mu =\frac{\nabla^\mu\phi}{(\nabla\phi)^2}$. 

Moreover the model (\ref{lag}) has the following symmetry 
\be 
\delta_E \phi=0 
\quad , \quad 
\delta_E g_{\mu\nu}= 
g_{\mu\nu}a_\sigma\nabla^\sigma\phi-\frac12 
\left( a_\mu \nabla_\nu\phi + a_\nu \nabla_\mu \phi \right)
\>, \label{deltaE}
\ee 
for arbitrary constant vector $a_\mu$.  
By means of the Noether theorem 
this symmetry can be easily shown  
to give rise to the following conserved current 
\be 
J^{\mu\nu}= g^{\mu\nu}E \label{E1}
\>, \ee
where 
\be 
E=\frac12\left[(\nabla\phi)^2 -
J(\phi)\right]
\>,
\ee
and
\be
J^{\prime}\left(\phi\right)=V\left(\phi\right)
\>.\label{J} 
\ee
The conservation law for $J^{\mu\nu}$ implies the space-time 
independence of the local energy $E$ 
\cite{Gegenberg}.  

In the present case, as in the CGHS  model, the conservation 
law for the currents $j_1^\mu$, $j_2^\mu$ turns out to imply the 
existence of two free fields. It is not difficult to check that 
$j^\mu_1$ and $j^\mu_2-j^\mu_R$ satisfy the integrability 
condition. The corresponding free-fields equations are: 
\bea 
\square j_1&=&0 \>, \nonumber\\ 
R +\square j_2 &=&0 \>,\label{ffields}
\eea 
where 
\be
j_1=\int^\phi {d\tau\over 2E+J\left(\tau\right)}\>,
\ee
and 
\be 
j_2 =\log(2E+J)=\log(\nabla\phi)^2 
\>. 
\ee 
We have to note that in the integral of (5.17) $E$ should be considered as a constant.
At this point it is clear that 
we have generalized the conformal symmetry of the CGHS model in 
the sense that it could be recovered as a special 
linear combination of $\delta_1$ and $\delta_2$.  
In section 3 we also showed that the 
CGHS model can be seen as a particular case ($\beta=0$) 
of a family of models (with potential 
$V=4\lambda^2\hbox{e}^{\beta\phi}$) having  
a conformal symmetry $\delta_\beta$. 
However, it is easy to see that, for non-constant potentials, 
no non-trivial linear combination of $\delta_1$ and $\delta_2$ 
is conformal. This suggests that it may exist another 
symmetry $\delta_3$, which must be independent 
from $\delta_1$ and $\delta_2$, such that a linear combination 
of  $\delta_1,\> \delta_2$ and $\delta_3$ gives rise to a 
conformal symmetry, at least in the particular case in which the 
potential is an exponential of the dilaton. 

To find out this symmetry we note that, for the 
exponential models, the Noether current  
associated to the conformal symmetry $\delta_\beta$ 
can be written as: 
\be 
j^\mu_\beta =j^{\mu}_2+2\beta Ej^\mu_1 \>.\label{jbeta}
\ee
However, since $E$ is constant, this current is also conserved for 
a generic 2D dilaton gravity model. Our task, therefore, is to 
work out, for a generic model, 
the Noether symmetry $\delta_3$ which 
is associated with the conserved 
current $Ej_1$. A straightforward application of the Noether 
theorem yields
\be 
\delta_3\phi=0 
\quad,\quad 
\delta_3 g_{\mu\nu}= -\frac{\epsilon_3}2\left[g_{\mu\nu}+ 
J\left({g_{\mu\nu}\over\left(\nabla\phi\right)^2}
-2{\nabla_{\mu}\phi
\nabla_{\nu}\phi\over\left(\nabla\phi\right)^4}\right)\right]
\>. \label{delta3}
\ee
Hence the conformal symmetry of the exponential models, for which 
$V=\beta J$, is given by 
\be 
\delta_\beta = \delta_2 +2\beta\delta_3
\>. \label{deltabeta}
\ee
Therefore, the CGHS model and the models with an 
exponential potential are special only in the sense that 

a) One of the symmetries $\delta_1,\>\delta_2$ and 
$\delta_3$ is a linear combination of the other two. 

b) A linear combination of $\delta_1,\>\delta_2$ and 
$\delta_3$ is conformal and does not 
involve the space-time derivatives of the fields. 

Finally, we mention that the three symmetries close down to a non-abelian
Lie algebra. The symmetry $\delta_2$ is a central generator, but $\delta_1$
and $\delta_3$ generate the affine algebra:
$[ \delta_1, \delta_3 ] = {1\over2} \, \delta_1$.

\subsection*{Construction of an invariant semiclassical action}

To complete the program which introduced the present section, we must
consider now the construction of an invariant metric $\bar{g}_{\mu\nu}$.
Although for a generic 2D dilaton gravity model the invariant metric will
not be unique, in the present paper, and for the sake of clarity, we shall
consider the simplest choice.

Here we shall consider the metric $\bar{g}_{\mu\nu}$ which fulfils the 
following requirements: 

a) It is invariant under $\delta =\delta_2-4\lambda^2\delta_1$,  

b)  $\bar g_{\mu\nu}\equiv g_{\mu\nu}$ when $V=4\lambda^2$, and  

c)  det$\ g_{\mu\nu}$ $=$ det$\ \bar g_{\mu\nu}\ $. 

The requirements a) and b) guarantee that, when $V=4\lambda^2$, 
our semiclassical model will reduce to the BPP model 
(or RST model). The requirement c) appears to be a natural 
one since the symmetry $\delta$ is area preserving. 

Eqs. (\ref{hv}) and (\ref{hvii}) 
suggest that an invariant metric may be of the form: 
\be 
\bar g_{\mu\nu}=\tilde{A} g_{\mu\nu}+
\tilde{B}\nabla_{\mu}\phi\nabla_{\nu}\phi
\>,\label{hxi}
\ee
where $\tilde{A}=\tilde{A}(g_{\mu\nu},\phi),
\> \tilde{B}=\tilde{B}(g_{\mu\nu}, \phi)$ are 
scalar functions to be determined. It can be 
written in the form
\be
 \bar g_{\mu\nu}=A \left(\frac{g_{\mu\nu}}{(\nabla\phi)^2}-
\frac{\nabla_{\mu}\phi\nabla_{\nu}\phi}{(\nabla\phi)^4}\right) 
+B\nabla_{\mu}\phi\nabla_{\nu}\phi\>,\label{hxia}
\ee
where the new scalars $A=A(g_{\mu\nu}, \phi)$ 
and $B=B(g_{\mu\nu}, \phi)$ multiply quantities which are   
invariant under $\delta$. Therefore $A$ and $B$ must also be 
invariant. The simplest scalar which is invariant under 
$\delta$ is 
\be 
E_\lambda = \frac12\left((\nabla\phi)^2 -J(\phi)  
+4\lambda^2\phi\right)\equiv E+2\lambda^2\phi
\>.\label{Elambda}
\ee
Therefore, it appears natural to consider that $A$ and $B$ 
are functions of $E_\lambda$. 

The condition c) implies $AB=1$. Moreover, since 
for $V=4\lambda^2$  we have 
$(\nabla\phi)^2=2E_\lambda$, condition b) requires 
\be 
A = 2E_\lambda \>.
\ee 
Therefore, a metric which fulfils the three requirements above 
is 
\be
\bar g_{\mu\nu} = 
\frac{2E_\lambda}{\left(\nabla\phi\right)^2}g_{\mu\nu}+
\left(\frac1{2E_\lambda}-{\frac{2E_\lambda} 
{\left(\nabla\phi\right)^4}}\right)
\nabla_{\mu}\phi\nabla_{\nu}\phi\>.\label{hxii}
\ee 
The inverse metric is given by 
\be
\bar g^{\mu\nu} = 
\frac{\left(\nabla\phi\right)^2}{2E_\lambda}g^{\mu\nu}+
\left({\frac{2E_\lambda} 
{\left(\nabla\phi\right)^4}}-\frac1{2E_\lambda}\right)
\nabla^{\mu}\phi\nabla^{\nu}\phi
\>.\label{inversabarg}
\ee 

Since $\left(\bar\nabla \phi\right)^2=2E_\lambda$   
the inverse relation of (\ref{hxii}) takes the form  
\be 
g_{\mu\nu} = 
\frac{2\bar{E}_\lambda}{\left(\bar\nabla\phi\right)^2}g_{\mu\nu}+
\left(2\bar E_\lambda-{\frac{2\bar E_\lambda} 
{\left(\bar \nabla\phi\right)^4}}\right)
\nabla_{\mu}\phi\nabla_{\nu}\phi
\>, \label{hxiia}
\ee 
where 
\be 
\bar{E}_\lambda = \frac12\left(\left(\bar\nabla \phi\right)^2 
+J(\phi)-4\lambda^2\phi\right)
\>.\label{barElambda}
\ee 
Therefore, the inverse transformation 
is obtained from the direct one by (essentially) 
changing the sign of the potential. 

Once introduced the metric $\bar{g}_{\mu\nu}$ (\ref{hxii}), which is invariant under
the symmetry (\ref{delta}), we can immediately construct a semiclassical
action which preserves this symmetry by conformally coupling the matter
fields to the metric $\bar{g}_{\mu\nu}$ and, using the inverse relation
(\ref{hxiia}), write the action in terms of the metric $\bar{g}_{\mu\nu}$:
\be
S = S_{DG}\left[g(\bar{g},\phi),\phi\right]
- {1\over2} \sum_{i=1}^N \int d^2 x \sqrt{-\bar{g}} \bar{g}^{\mu\nu}
\partial_\mu f_i \partial_\nu f_i 
+ S_P (\bar{g})
\>, \label{twi}
\ee
where $S_{DG}$ is the dilaton-gravity sector of the action $S_0$.
Such an action will obviously reduce to the BPP model
(\ref{bppi}), (\ref{bppii}), for $V = 4\lambda^2$.
In conclusion, we have provided a semiclassical action invariant under the
transformation
\bea
\delta \phi &=& \epsilon \>, \nonumber \\
\delta \bar{g}_{\mu\nu} &=& 0 \>. 
\eea
This is the standard expression of the symmetry of the CGHS and exponential
model, which allows to reduce the associated sigma model to a Liouville-type
theory.
We expect that a generalization of the approach of Ref.~\cite{Kazama} to a
second order non-linear sigma model could imply solubility of the theory, 
in terms of the invariant metric.
This question will be considered elsewhere \cite{inprogress}.

\section{Conclusions and final comments}

In this paper we have considered the problem of generalizing the extra symmetry of the CGHS model.
This symmetry plays a crucial role in the semiclassical treatment of the CGHS theory and therefore it seems 
natural to look for some sort of generalization.
Our approach has two folds.

First of all, the conformal nature of the CGHS extra symmetry suggests to use 
a sigma model formalism to classify the possible models possessing a 
conformal-type symmetry transformation.
Taking into account the fact that a generic 2D dilaton-gravity model can be brought to a form with a
vanishing kinetic term, the condition for having an extra conformal symmetry of 
Ref.~\cite{Kazama} can be solved immediately.
We have found that the unique models admitting an extra conformal symmetry are those with an exponential
potential $V=4\lambda^2e^{\beta\phi}$.
Therefore, any theory obtained from the model with an exponential potential by
(conformal) redefinitions of the fields inherits an extra conformal symmetry.
For $\beta= 0$ an elementary change of variables transforms the reduced 
action into the CGHS standard action, and a further field redefinition of 
the fields generates also the models of
\cite{Fabbri-Russo}. 
When $\beta\neq 0$ the reduced form of the action itself leads to black hole
  solutions with Hawking temperature proportional to the mass.
The analogue of the BPP model can be constructed at once by adding a Polyakov-Liouville term respect
to the invariant metric.
The semiclassical theory could be exactly solved for especial forms of the energy-momentum tensor
$T^f_{\pm\pm}$ and the boundary conditions  $t_{\pm}$. 
In general, the problem reduces to solve Ricatti differential equations.
If $T_{\pm\pm}^f=0$ and $t_{\pm}=0$, 
the solution describes a black hole in thermal equilibrium and an 
adequate modification of the boundary conditions, to decrease the
incoming thermal flux, can describe an 
evaporating black hole with a non-constant temperature \cite{Bath}. 

Secondly, we have introduced three different extra symmetries for a generic
2D dilaton gravity model. These symmetries, which in general are
non-con\-formal, turn out to be conformal for particular linear combination of
them and reduce to the symmetry of the exponential model. This intriguing 
relation between the symmetries suggests that a special non-conformal symmetry
could play the same role as the standard conformal symmetry of the CGHS model
for a generic 2D dilaton gravity theory.  
This could open an avenue to study the semiclassical evolution 
of Schwarzschild black holes in an analytical setting.
\section{Acknowledgements}

J.~N-S. would like to thank J.~M.~Izquierdo and A. Mikovic for interesting discussions.
M.~N. is grateful to the Spanish MEC, CSIC and also the IMAFF for a research 
contract.


\begin{thebibliography}{99}

\bibitem{CGHS}
	C. G. Callan, S. B. Giddings, J. A. Harvey and A. Strominger,
	{\it Phys. Rev.}\/ D45 (1992) 1005.
\bibitem{Polyakov}
	A. M. Polyakov, {\it Phys. Lett.}\/ B103 (1981) 207.
\bibitem{RST}
	J. G. Russo, L. Susskind and L. Thorlacius, {\it Phys. Rev.}\/ 
	D46 (1993) 3444; {\it Phys. Rev.}\/ D47 (1993) 533.
\bibitem{Giddings}
	S. B. Giddings, {\it Quantum Mechanics of Black Holes}, 
	hep-th/9412138.
\bibitem{Strominger}
	A. Strominger, {\it Les Houches Lectures on Black Holes},
	hep-th/9501071. 
\bibitem{Thorlacius}
	L. Thorlacius, {\it Black Hole Evolution}, hep-th/9411020.
\bibitem{BPP}
	S.  Bose, L.  Parker and Y.  Peleg, 
	{\it Phys. Rev.}\/ D 52 (1995) 3512; 
	{\it Phys. Rev. Lett.}\/ 76 (1996) 861; 
	{\it Phys. Rev.} D 53 (1996) 7089.  
\bibitem{Cruz}
	J.~Cruz and J.~Navarro-Salas, {\it Phys.~Lett.} B 375 (1996) 47
\bibitem{Kazama}
	Y. Kazama, Y. Satoh and A. Tsuchiya, {\it Phys. Rev.} D51 (1995) 4265.
\bibitem{Bilal}
	A. Bilal and C. G. Callan, {\it Nucl. Phys.}\/ B394 (1993) 73. \\
	S. de Alwis, {\it Phys. Lett.}\/ B289 (1992) 278
\bibitem{Mikovic} 
        A. Mikovic, {\it Class. Quant. Grav.} 13 (1996) 209;
	{\it Phys. Lett.}\/ B355 (1995) 85.
\bibitem{Michaud}	
	G.  Michaud and R.  C.  Myers, {\it Two dimensional Dilaton Black Holes},
	gr-qc/9508063.
\bibitem{Jackiw}
	R.~Jackiw, {\it Another view on Massless Matter-Gravity Fields in
	Two Dimensions}, hep-th/9501016.
\bibitem{Navarro}
	J.  Navarro-Salas, M.  Navarro and C.  F.  Talavera, 
	{\it Phys. Lett.}\/ B 356 (1995) 217.
\bibitem{Banks}
        T.  Banks and M.  O'Loughlin, {\it Nucl. Phys.} B362 (1991) 649.
\bibitem{Gegenberg}
	J.~Gegenberg, G.~Kunstatter and D.~Louis-Martines, 
	{\it Phys. Lett.} B321 (1994) 193; {\it Phys. Rev.} D51 (1995) 1781;
	{\it Classical and Quantum Mechanics of Black Holes in Generic 2D
	Dilaton Gravity}, gr-qc/9501017.
\bibitem{JT}
        R.  Jackiw, in {\it Quantum Theory of Gravity}, ed. S.  Christensen (Adam Hilger, Bristol, 1984) p. 403;
        C.  Teltelboim, in {\it Quantum Theory of Gravity}, ed. S.  Christensen (Adam Hilger, Bristol, 1984) p. 327.
\bibitem{Mann1}
         R.  B.  Mann, A.  Shiekh and L.  Tarasov, {\it Nucl. Phys.} B
         341(1990) 134-154.
\bibitem{Mann2}
R.  B.  Mann, {\it Nucl. Phys.} B 418(1994) 231-256.
\bibitem{inprogress}
	J. Cruz, J. Navarro-Salas, M. Navarro and C. F. Talavera, work in
	progress. 
\bibitem{Fabbri-Russo}
	A.  Fabbri and J.  G.  Russo, {\it Solvable models in 2D
	dilaton gravity}, CERN-TH/95-267, hep-th/9510109.
 \bibitem{Bath}
         J.  Cruz and J.  Navarro-Salas, 
          {\it Phys. Lett. }B 387(1996) 51 ;
          J.  Cruz, A.  Mikovic and J.  Navarro-Salas, {\it A Quantum Model of 
          Schwarzschild Black Hole Evaporation} hep-th/9611219 ({\it Phys.Lett. B}, to appear).  
\end{thebibliography}
\end{document}